\documentclass[10pt,letterpaper]{article}
\usepackage{opex3}

\usepackage{cite}
\usepackage{amsmath,amssymb,latexsym,float,epsfig,subfigure,textcomp}
\DeclareMathOperator{\um}{\mu m}
\DeclareMathOperator{\nm}{nm}

\begin{document}

\title{Ultra-compact On-Chip Plasmonic Light Concentrator}

\vskip4pc


\title{Ultra-compact on-chip plasmonic light concentration based on a hybrid photonic-plasmonic structure}

\author{Ye Luo, Maysamreza Chamanzar and Ali Adibi}

\address{School of Electrical and Computer Engineering, Georgia Institute of Technology,  \\ 777 Atlantic Drive NW, \\ Atlanta, GA 30332--0250}

\email{luoye@ece.gatech.edu}

\begin{abstract}
We present a novel approach for achieving tightly concentrated optical field by a hybrid photonic-plasmonic device in an integrated platform, which is a triangle-shaped metal taper mounted on top of a dielectric waveguide. This device, which we call a plasmomic light concentrator (PLC), can achieve side-coupling of light energy from the dielectric waveguide to the plasmonic region and light focusing into the apex of the metal taper(at the scale $\sim10\nm$) at the same time. For demonstration, we numerically investigate a PLC, which is a metal (Au) taper on a dielectric ($\text{Si}_3\text{N}_4$) waveguide at working wavelengths around $800\nm$. We show that three major effects (mode beat, nanofocusing, and weak resonance) interplay to generate this light concentration phenomenon and govern the performance of the device. By coordinating these effects, the PLC can be designed to be super compact while maintaining high efficiency over a wide band. In particular, we demonstrate that under optimized size parameters and wavelength a field concentration factor (FCF), which is the ratio of the norm of the electric field at the apex over the average norm of the electric field in the inputting waveguide, of about $13$ can be achieved with the length of the device less than $1\um$ for a moderate tip radius $20\nm$. Moreover, we show that a FCF of $5-10$ is achievable over a wavelength range $700-1100\nm$ when the length of the device is further reduced to about $400\nm$.
\end{abstract}



\section{Introduction}
Plasmonic metallic structures are well-known for the ability of breaking the diffraction limit and achieving large field enhancements or field confinements in small volumes\cite{Schuller10}. One mainstream of the approaches is to focus surface plasmon polariton (SPP) modes, which can be further divided into two categories. One is interference-induced plasmonic nanofocusing. For this type of techniques, the plasmonic excitation sites on a metal film are carefully tuned so that the propogation of SPP originated at the sites can get focused into a hot spot on the metal film. Typical planar patterns for plasmonic excitation on the metal film include individual circular rings~\cite{LSSPSZ05}, multiple concentric rings~\cite{CANZ09}, curved arrays of nanoholes~\cite{Yin05}, and certain coaxial structures~\cite{YL09}. The other is geometry-induced plasmonic nanofocusing. In principle, tapering the cross section of the SPP propagation gradually will increase the intensity of the SPP and the slow down its group velocity. At the tip end of the taper, the group velocity becomes infinity which means the surface plasmon becomes localized and the energy is highly concentrated~\cite{Nerkararyan97,BMN00,Stockman04,VGP07,Moreno09,BN09,Volkov09,Volkov09:2,FQWZ10,TKT10,Lindquist10,Ropers07,VPK08,Normatov10,Choi09,Schnell11}. This type of techniques can accurately control the focus point and usually with a higher concentrating ability.

Conventionally, the nanofocusing techniques have the SPP excited from free space and the devices are usually bulky. There are two issues which limit the concentration efficiency. First, the efficiency of SPP excitation is usually low and unstable. Second, the SPP propagation length is limited by the lossy nature of metal. Very recently, several approaches of realizing plasmonic nanofocusing on integrated-optics platforms have been reported~\cite{YWLA08,DGL10,DGL09,DGL11,LCEA11,HYY11}. Using these on-chip light concentration techniques, the light energy is coupled from a photonic waveguide to a plasmonic structure (usually a metallic taper) and then concentrating the light energy to the tip of the taper. The coupling from the photonic waveguide to the plasmonic structure is efficient and robust, while the plasmonic part is usually compact since it only concentrates light energy from a ``$\sim\lambda$'' scale to a ``$<\lambda/10$'' scale. Therefore, the two limitations for free space plasmonic nanofocusing can be well addressed by the on-chip techniques. However, instead of demonstrating the promising plasmonic concentration phenomena, these reports lack a systematic analysis of the underlying mechanisms that governing the performance of the devices.

In this paper, we will present the design of a novel ultra-compact on-chip plasmonic light concentrator (PLC), and show that there are three major effects (mode beat, nanofocusing, and weak resonance) that govern this light concentration phenomenon and affect the efficiency of the device. By coordinating these effects, this proposed structure can maintain very high concentrating efficiency while the size dimensions (width$\sim400\nm$ length$<1\um$) are the smallest among the all the plasmonic-light-concentration techniques that have been reported.

\begin{figure}[htbp]
\centering
\includegraphics[width=.75\textwidth]{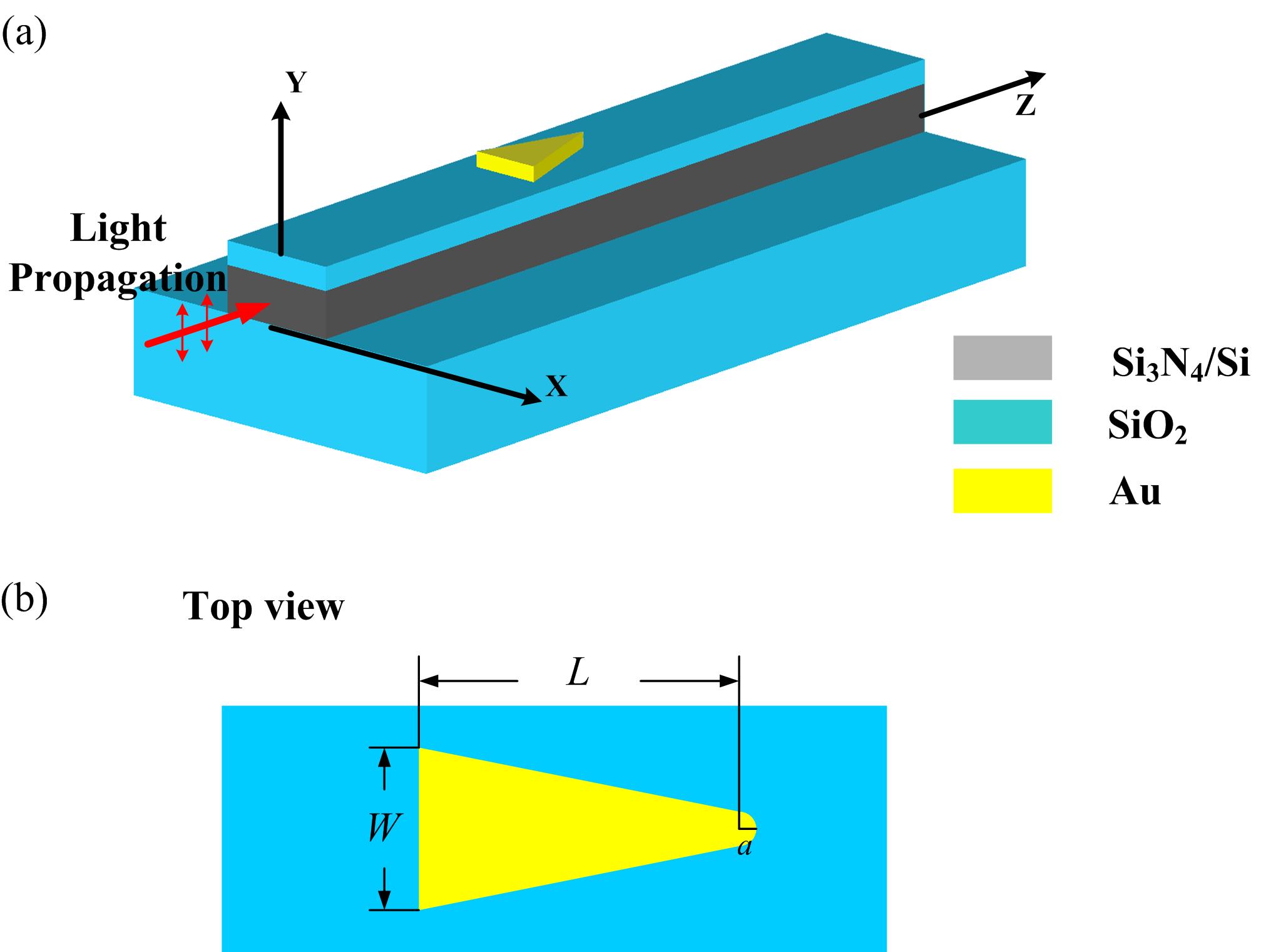}
\caption{(a) Schematic of an ultra-compact PLC that is a hybrid photonic-plasmonic structure with a gold triangle taper integrated on top of a dielectric ($\text{Si}_3\text{N}_4$ or Si) ridge waveguide with a $\text{SiO}_2$ buffer layer. (b) The top view of this hybrid structure.}\label{Fig:PLC:Schematic}
\end{figure}

The basic structure of the device is a small triangle-shaped metal taper on top of a dielectric waveguide, as shown in Figure~\ref{Fig:PLC:Schematic}(a). In our analysis of this structure, the material for the metal triangle is gold (Au), and the material for the dielectric ridge waveguide is silicon nitride ($\text{Si}_3\text{N}_4$). The substrate is silicon dioxide ($\text{SiO}_2$). In addition, a $\text{SiO}_2$ buffer layer is used to separate the metal triangle and the dielectric ridge waveguide. The ambient material is water. The dimensions of the ridge waveguide are chosen to support only the fundamental TM-like (vertical polarization) mode and the fundamental TE-like (horizontal polarization) mode. Only the fundamental $\text{TM}$-like mode can be used to generate light concentration. We set up the coordinate system with $X$-axis in the horizontal direction, $Y$-axis in the vertical direction and $Z$-axis in the propagation direction of the input light.

Figure~\ref{Fig:PLC:Schematic}(b) shows the top view of the metal triangle, which is isosceles. We use $W$ to denote the length of the base of the isosceles triangle, and $L$ to denote the the length of a perpendicular from the center of the curved tip to the middle point of the base side. In particular, we will call $W$ the width of the triangle and $L$ the length of the triangle throughout this paper. Since the apex of the metal triangle is always rounded up in real fabrications, we introduce the radius of curvature $a$ at the tip as a parameter.


\section{Mode analysis of the hybrid photonic-plasmonic waveguide} \label{S:PLC:HybridWG}
\begin{figure}[htbp]
\centering
\includegraphics[width=\textwidth]{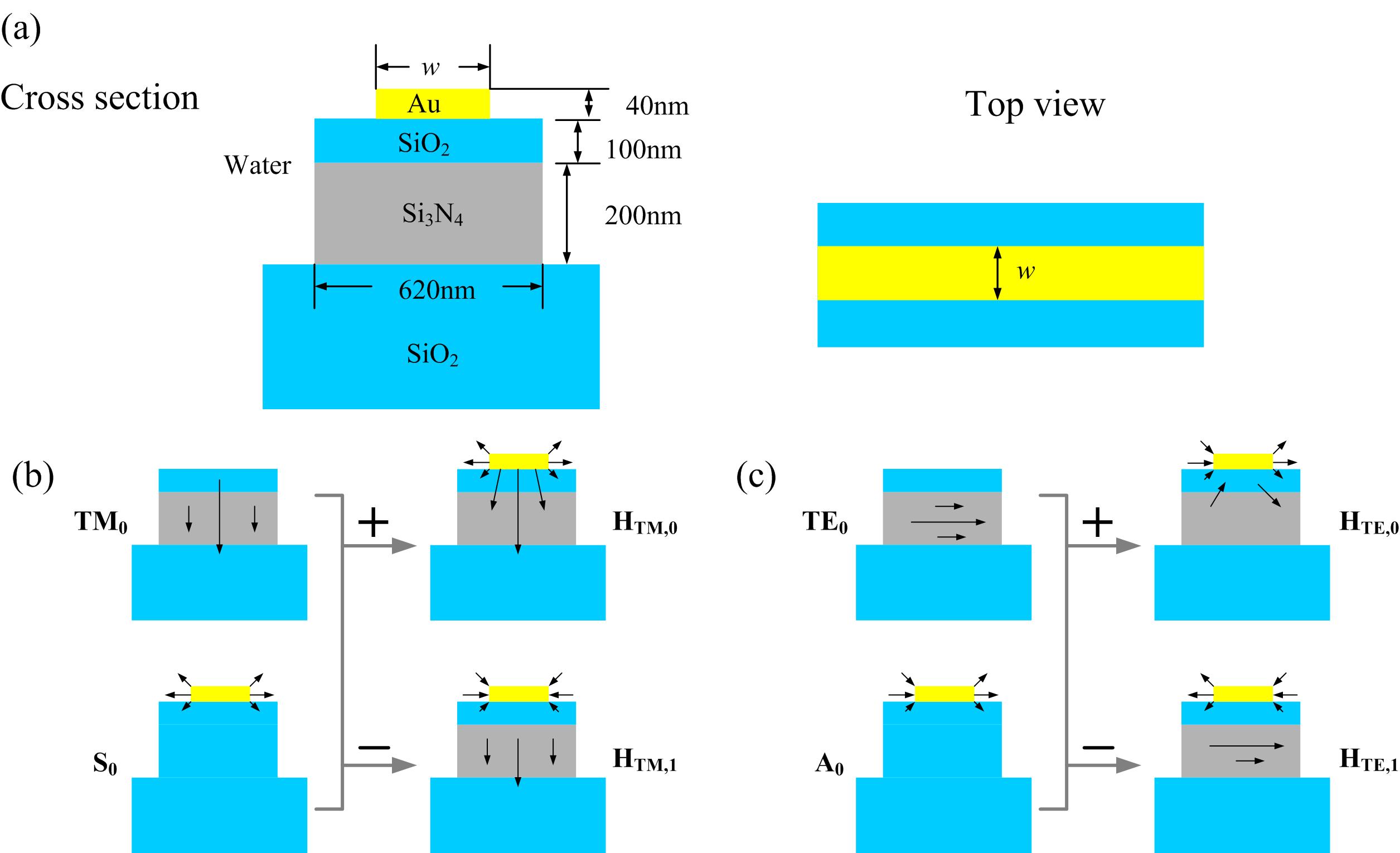}
\caption{(a) The cross section and top view of a hybrid photonic-plasmonic waveguide. (b) Two supermodes ($\text{H}_{\text{TM},0}$ and $\text{H}_{\text{TM},1}$) come from the superposition of the fundamental TM-like mode ($\text{TM}_0$) of the purely photonic waveguide and the fundamental symmetric mode ($\text{S}_0$) of the purely plasmonic waveguide. (c) Two supermodes ($\text{H}_{\text{TE},0}$ and $\text{H}_{\text{TE},1}$) come from the superposition of the fundamental TE-like mode ($\text{TE}_0$) of the purely photonic waveguide and the fundamental asymmetric mode ($\text{A}_0$) of the purely plasmonic waveguide.
}\label{Fig:PLC:ModeSplit}
\end{figure}

To analyze the behavior of the PLC, we first do mode analysis on a hybrid photonic-plamsmonic waveguide, which is a metal strip on top of a dielectric ridge waveguide with a buffer layer. (There have been reported hybrid photonic-plamsmonic waveguides with different geometric settings. See \cite{Berini00,Berini01,CBBL00,Park10,GDL10:2,DH10}.) As shown in Figure~\ref{Fig:PLC:ModeSplit}(a), the $\text{Si}_3\text{N}_4$ ridge waveguide has dimensions of $620\nm$ in width and $200\nm$ in thickness. The thickness of the $\text{SiO}_2$ buffer layer is $100\nm$. The Au strip has thickness of $40\nm$ and width of $w$. The ambient material is water. Assume the working wavelength $\lambda$ is $800\nm$. At this wavelength, the refractive index of $\text{Si}_3\text{N}_4$ is assumed to be $2$, the refractive index of $\text{SiO}_2$ is assumed to be $1.46$, the refractive index of water is assumed to be $1.33$, and the permittivity of Au is assumed to be $-24.02+j1.18$.

Here we study the supermodes in the hybrid waveguides by looking at possible mode couplings between the modes in the corresponding purely photonic waveguide (with no Au layer) and the modes in the corresponding purely plasmonic waveguide (having $\text{Si}_3\text{N}_4$ changed to $\text{SiO}_2$). Under the aforementioned dimensions, the purely photonic waveguide only support the fundamental TM-like mode (denoted by $\text{TM}_0$) and the fundamental TE-like mode (denoted by $\text{TE}_0$). The $\text{TM}_0$ mode can only couple with symmetric modes of the purely plasmonic waveguide, and the $\text{TE}_0$ mode can only couple with the asymmetric modes of the purely plasmonic waveguide. We only consider the fundamental symmetric ($\text{S}_0$) and asymmetric ($\text{A}_0$) modes of the purely plasmonic waveguide, since higher order modes are cut off with the dimensions of our structure at the wavelength $\lambda$ of $800\nm$. Figure~\ref{Fig:PLC:ModeSplit}(b) shows two supermodes, $\text{H}_{\text{TM},0}$ and $\text{H}_{\text{TM},1}$, are derived from the superposition of the $\text{TM}_0$ mode and the $\text{S}_0$ mode. The electric field lines are sketched for these modes. Analogously, Figure~\ref{Fig:PLC:ModeSplit}(c) shows two supermodes, $\text{H}_{\text{TE},0}$ and $\text{H}_{\text{TE},1}$, are derived from the superposition of the $\text{TE}_0$ mode and the $\text{A}_0$ mode.

To give more details about the mode-coupling processes, we use finite element method (FEM, COMSOL Multiphysics) to compute the eigenmodes at the wavelength $\lambda$ of $800\nm$. Note that the effective indices $n_\text{eff}$ of the propagating modes must have real parts greater than $1.46$, which is the refractive index of the $\text{SiO}_2$ substrate.

\begin{figure}[htbp]
\centering
\includegraphics[width=0.75\textwidth]{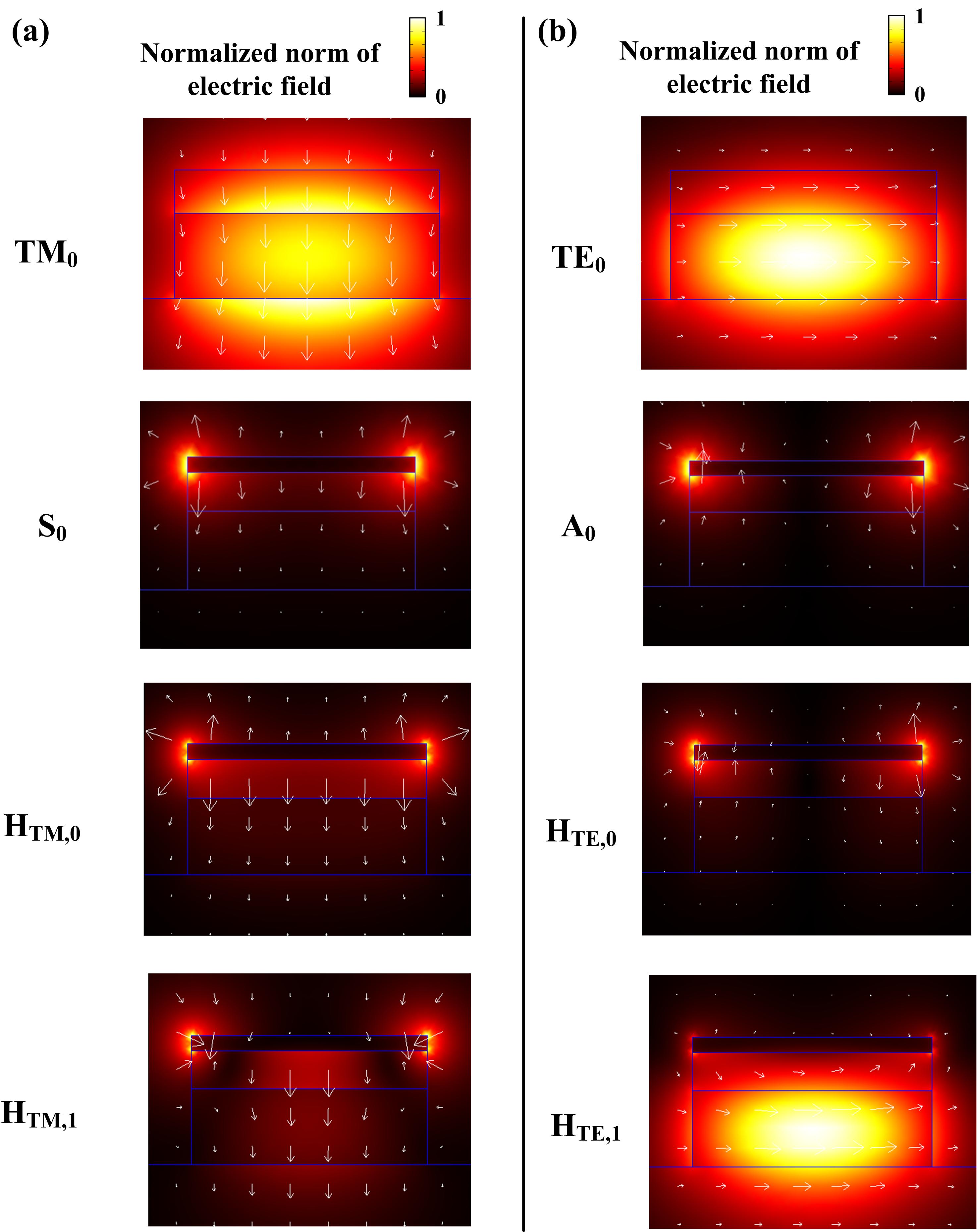}
\caption{(a) The normalized electric field profiles of $\text{TM}_0$, $\text{S}_0$, $\text{H}_{\text{TM},0}$ and $\text{H}_{\text{TM},1}$. (b) The normalized electric field profiles of $\text{TE}_0$, $\text{A}_0$, $\text{H}_{\text{TE},0}$ and $\text{H}_{\text{TE},1}$. The width $w$ of the Au layer is $620\nm$. The wavelength $\lambda$ is $800\nm$.}\label{Fig:PLC:FieldSiN_TM_TE}
\end{figure}

Figure~\ref{Fig:PLC:FieldSiN_TM_TE}(a) shows the normalized electric field profiles of $\text{TM}_0$, $\text{S}_0$, $\text{H}_{\text{TM},0}$ and $\text{H}_{\text{TM},1}$ modes, with the width $w$ of the Au strip set to $620\nm$. The electric field lines are plotted. This corresponds to the mode coupling process explained in Figure~\ref{Fig:PLC:ModeSplit}(b). The $\text{TM}_0$ mode has an effective index of $n_\text{eff}=1.5696$, the $\text{S}_0$ mode has an effective index of $n_\text{eff}=1.7302+j0.0223$, the $\text{H}_{\text{TM},0}$ mode has an effective index of $n_\text{eff}=1.5743+j0.0094$, and the $\text{H}_{\text{TM},1}$ mode has an effective index of $n_\text{eff}=1.7992+j0.0163$. Analogously, the related modes in Figure~\ref{Fig:PLC:ModeSplit}(c), $\text{TE}_0$, $\text{A}_0$, $\text{H}_{\text{TE},0}$ and $\text{H}_{\text{TE},1}$, are illustrated in Figure~\ref{Fig:PLC:FieldSiN_TM_TE}(b). The $\text{TE}_0$ mode has an effective index of $n_\text{eff}=1.6533$, the $\text{A}_0$ mode has an effective index of $n_\text{eff}=1.7088+j0.0238$, the $\text{H}_{\text{TE},0}$ mode has an effective index of $n_\text{eff}=1.7512+j0.0213$, and the $\text{H}_{\text{TE},1}$ mode has an effective index of $n_\text{eff}=1.6273+j6.1\times10^{-4}$.

\begin{figure}[htbp]
\centering
\includegraphics[width=1.0\textwidth]{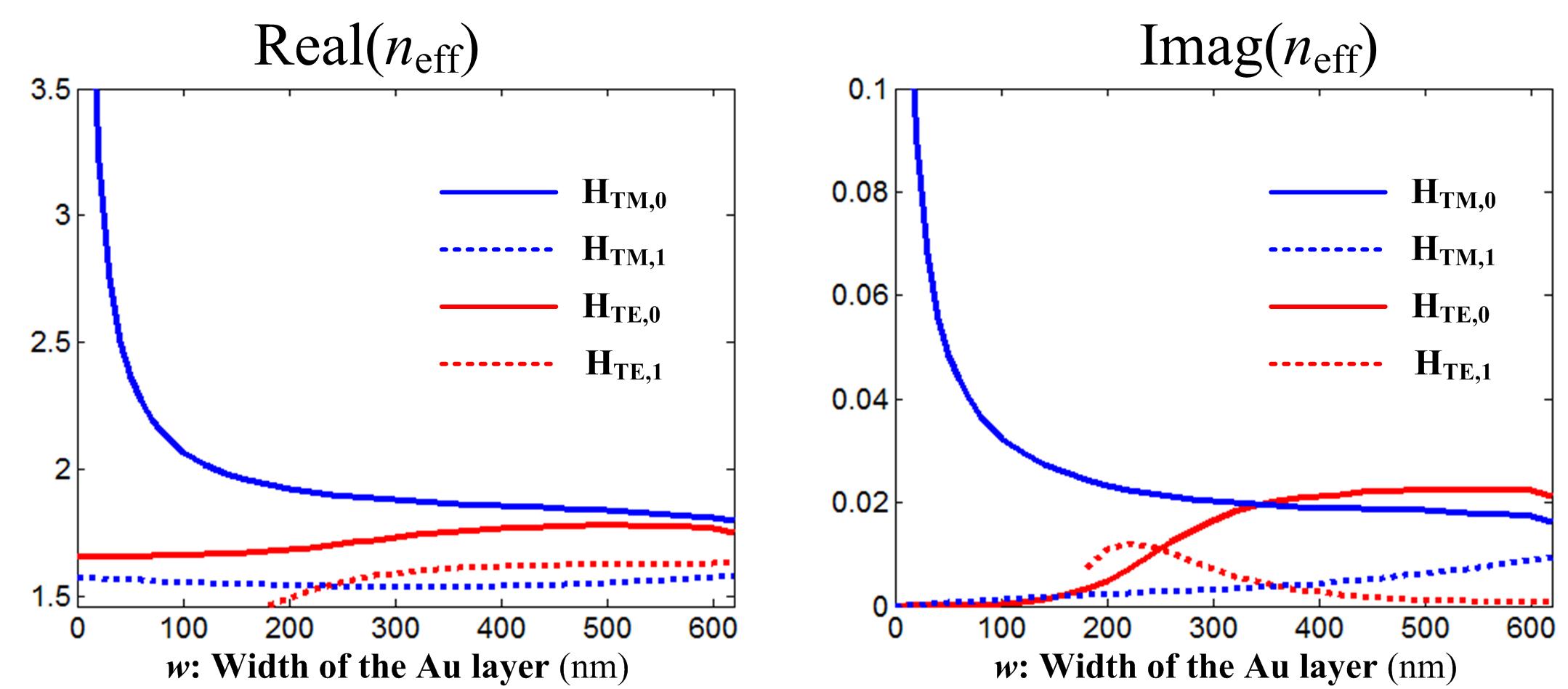}
\caption{Dispersion characterization of the four supermodes ($\text{H}_{\text{TM},0}$, $\text{H}_{\text{TM},1}$, $\text{H}_{\text{TE},0}$ and $\text{H}_{\text{TE},1}$) with respect to the width $w$ of the Au strip. The wavelength $\lambda$ is $800\nm$.}\label{Fig:PLC:DispersionSiN}
\end{figure}

The dispersions of the four propagating supermodes ($\text{H}_{\text{TM},0}$, $\text{H}_{\text{TM},1}$, $\text{H}_{\text{TE},0}$ and $\text{H}_{\text{TE},1}$) are plotted in Figure~\ref{Fig:PLC:DispersionSiN} as the real and imaginary parts of the effective index versus the width $w$ of the Au strip. The coordinate ``$\text{Real}(n_\text{eff})$'' has minimum $1.46$, which is the refractive index of the $\text{SiO}_2$ substrate. As we have mentioned, the real part of the effective index of a propagating mode must be greater than $1.46$. The imaginary part of the effective index accounts for the propagation loss.

\begin{figure}[htbp]
\centering
\includegraphics[width=0.8\textwidth]{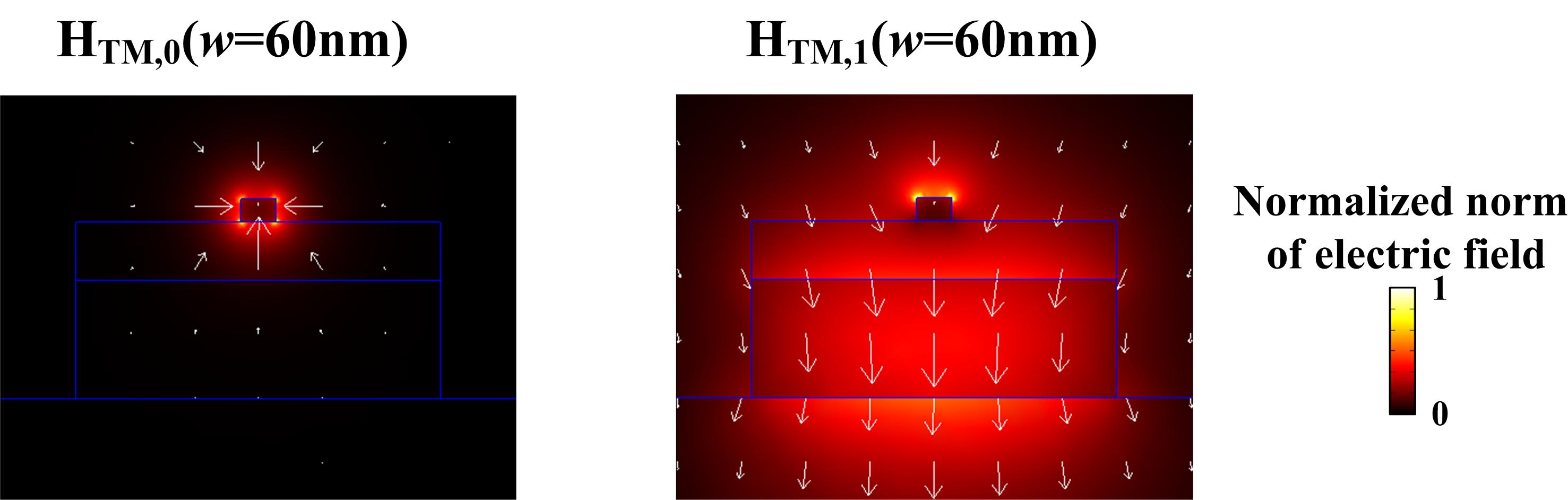}
\caption{The normalized electric field profiles of $\text{H}_{\text{TM},0}$ and $\text{H}_{\text{TM},1}$. The width $w$ of the Au strip is $60\nm$. The wavelength $\lambda$ is $800\nm$.}\label{Fig:PLC:FieldSiN_AuWidth60nm}
\end{figure}

The $\text{Real}(n_\text{eff})$ of $\text{H}_{\text{TM},0}$ is greater than the $\text{Real}(n_\text{eff})$ of $\text{H}_{\text{TM},1}$, and as $w$ decreases, the difference becomes larger. The imaginary parts follow the same trend. This is because as $w$ decreases, the light energy becomes more concentrated in the lossy metal region for $\text{H}_{\text{TM},0}$, while the light energy becomes more concentrated in the dielectric region for $\text{H}_{\text{TM},1}$. The mode profiles of both supermodes are illustrated in Figure~\ref{Fig:PLC:FieldSiN_AuWidth60nm} for $w=60\nm$. As $w$ approaches $0$, both the real and imaginary parts of the effective index of $\text{H}_{\text{TM},0}$ goes to infinity, which means that the phase velocity goes to $0$ and $\text{H}_{\text{TM},0}$ becomes more and more like a localized mode. When $w=0$, there is no metal strip and the $\text{H}_{\text{TM},1}$ mode is actually the purely photonic mode $\text{TM}_0$. Note that as the width of the metal layer continuously and gradually shrinks down, the amplitude of the $\text{H}_{\text{TM},0}$ mode increases since its propagating velocity slows down. Such a process is common to most geometry-induced plasmonic nanofocusing techniques, and is one of the main mechanisms to generate light concentration in our triangle-shaped plasmonic device.
Analogously, we observe that the $\text{H}_{\text{TE},0}$ mode evolves into the purely photonic mode $\text{TE}_0$ as $w$ approaches $0$. Moreover, when $w<180\nm$, the $\text{H}_{\text{TE},1}$ mode is cut off. Therefore we can not take use of the supermodes $\text{H}_{\text{TE},0}$ and $\text{H}_{\text{TE},1}$ for our application of light concentration.

The supermodes in the hybrid waveguide can be excited by the modes in the corresponding bare (or purely photonic) waveguide. More specifically, the incident mode $\text{TM}_0$ in the bare waveguide can excite two supermodes ($\text{H}_{\text{TM},0}$ and $\text{H}_{\text{TM},1}$) of the hybrid waveguide. The superposition of these two supermodes at the input end has field in the dielectric region to match the incident $\text{TM}_0$ mode. Since they have different effective indices, as can be seen from the dispersion diagram in Figure~\ref{Fig:PLC:DispersionSiN}, the two supermodes undergo a beat effect with light energy bouncing back and forth between the dielectric region and the metal region.  For a specific width $w$ of the Au strip, we have
\begin{equation} \label{Eq:Beatlength}
    \text{BL}(w)=\frac{\lambda}{\text{Real}(n_\text{eff}(\text{H}_{\text{TM},0}(w))-n_\text{eff}(\text{H}_{\text{TM},1}(w)))},
\end{equation}
where $\text{BL}(w)$ is the beat length. In particular, at a distance about half the beat length from the input end, the light energy is side-coupled efficiently into the metal region.

\section{A PLC with specific size parameters} \label{S:PLC:SiN}
The PLC presented in this paper can be considered as a taper of the metal strip of the hybrid waveguide introduced in the previous section. The performance of a PLC depends on the interplay of three effects: mode beat, nanofocusing and weak resonance. More specifically, the mode beat effect comes from the interference of the two excited supermodes ($\text{H}_{\text{TM},0}$ and $\text{H}_{\text{TM},1}$), the nanofocusing effect comes from the triangle taper, and the weak resonance effect comes from the reflections at the input end and the apex of the taper. In addition, we may design the device to be ultra-compact while maintaining high efficiency under a wide-band operation.  We will demonstrate these effects by analyzing the following device with specific size parameters.

For the dielectric part, we use the same size parameters as in the previous section, i.e., the $\text{Si}_3\text{N}_4$ ridge of width $620\nm$ and thickness $200\nm$, and the $\text{Si}\text{O}_2$ buffer of width $620\nm$ and thickness $100\nm$. The thickness of the Au triangle is $40\nm$. The ambient material is also water. The coordinate system is set up as in Figure~\ref{Fig:PLC:Schematic}. In particular, the coordinate origin is chosen such that the plane $X=0$ cuts through the middle of the $\text{Si}_3\text{N}_4$ ridge vertically, the plane $Y=0$ coincides with the boundary between the $\text{Si}_3\text{N}_4$ layer and the $\text{SiO}_2$ substrate, and the plane $Z=0$ goes through the center of the curved tip.

\begin{figure}[htbp]
\centering
\includegraphics[width=.8\textwidth]{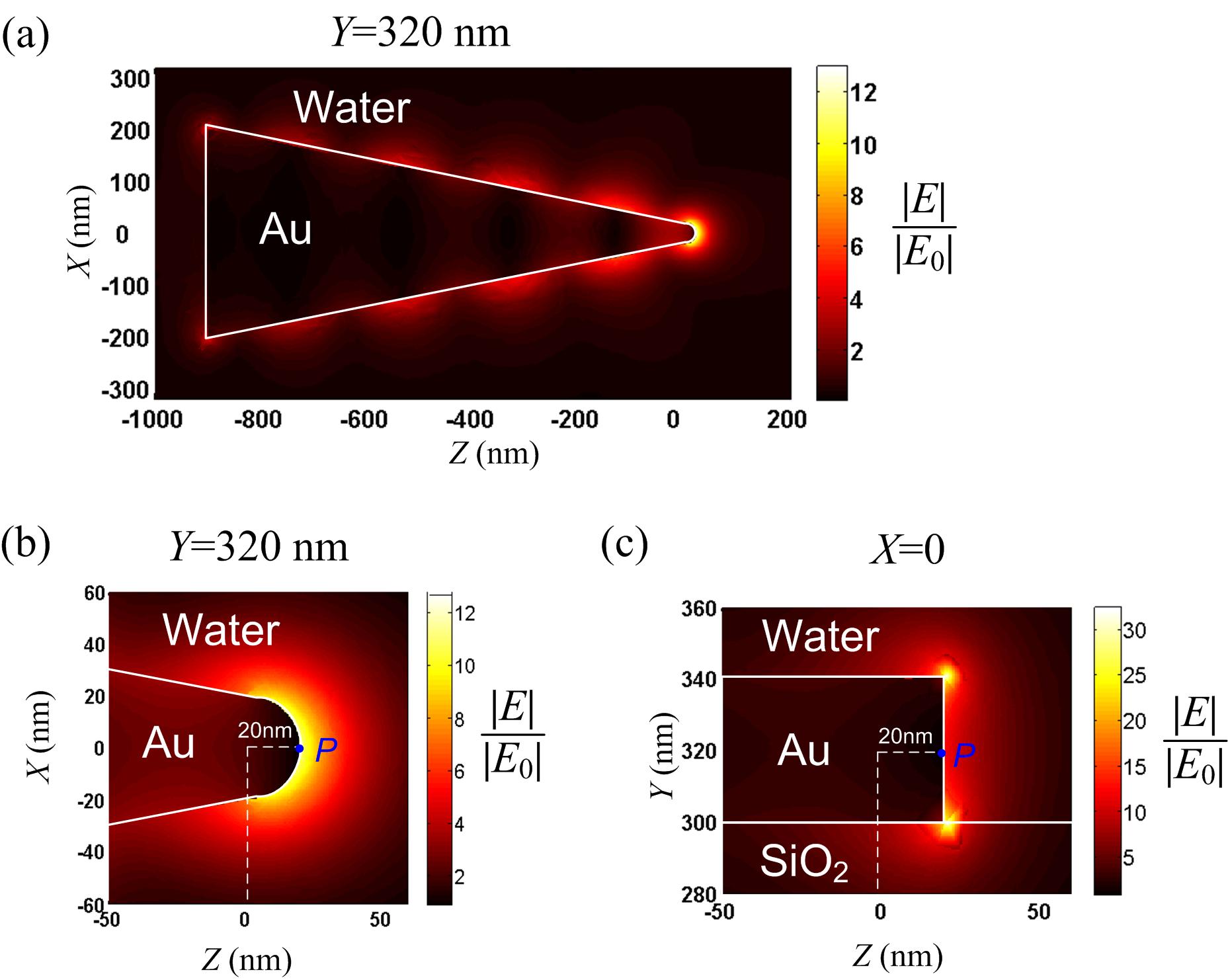}
\caption{Normalized electric field patterns in the planes horizontally ($Y=320\nm$) and vertically ($X=0$) cutting through the Au layer. The length $L$ of the Au triangle is $900\nm$, and the width $W$ is $400\nm$. The FCF is $12.6$ with he radius of curvature $a$ at the tip being $20\nm$.}\label{Fig:PLC:SiNFocus}
\end{figure}

As shown in Figure~\ref{Fig:PLC:SiNFocus}, we compute the electric field patterns in the horizontal ($Y=320\nm$ in (a) and (b)) and vertical planes ($X=0$ in (c)) using a $3$D FEM simulation (COMSOL Multiphysics). Here $Y=200\nm+100\nm+20\nm$, which is the sum of the thickness of the $\text{Si}_3\text{N}_4$ layer, the thickness of the $\text{SiO}_2$ buffer layer and half of the thickness of the Au layer. The Au triangle has dimensions of $L=900\nm$, $W=400\nm$ and $a=20\nm$. We let $|E_0|$ be the average norm of the incident electric field in the $\text{Si}_3\text{N}_4$ region, which is calculated from the the $\text{TM}_0$ mode of corresponding purely photonic waveguide. We estimate the total power flow by a plane-wave approximation with the electric field $E_0$ homogeneously distributed and restricted within the $\text{Si}_3\text{N}_4$ region. In particular, the norm of $E_0$ can be computed by the formula
\begin{equation}\label{Eq:AverageE1}
    |E_0| = \sqrt{\frac{2\eta_0}{n_{\text{Si}_3\text{N}_4}A_{\text{Si}_3\text{N}_4}}\int_D{\bar{\mathbf{S}}_z}}=\sqrt{\frac{2\eta_0}{n_{\text{Si}_3\text{N}_4}A_{\text{Si}_3\text{N}_4}}\int_D{\text{Re}(\frac{1}{2}(\mathbf{E}\times\mathbf{H^*})_z)}}.
\end{equation}
Here $\eta_0=\sqrt{\mu_0/\epsilon_0}\approx120\pi$ is the wave impedance of free space, $n_{\text{Si}_3\text{N}_4}=2$ is the refractive index of $\text{Si}_3\text{N}_4$, $A_{\text{Si}_3\text{N}_4}=620\nm\times200\nm$ is the area of the cross section of the $\text{Si}_3\text{N}_4$ slab, and $\int_D{\bar{\mathbf{S}}_z}$ is the time-averaged power flow through a cross-section plane $D$. The electric field patterns shown in Figure~\ref{Fig:PLC:SiNFocus} are normalized to $|E_0|$. The light field becomes highly concentrated at the metallic tip. To evaluate the efficiency of field concentration induced by the PLC, we introduce a field concentration factor (FCF) as the normalized electric field norm at the apex point (point $P$ in Figure~\ref{Fig:PLC:SiNFocus} (b) and (c)) in the horizontal plane cutting through the Au layer. In particular, the FCF is $12.6$ in this case. Note that the FCF depends on the radius of curvature $a$. While reducing $a$ can afford higher FCF, we should choose $a$ large enough for a good estimation on real fabrications. Here we set $a=20\nm$, and write $\text{FCF}(a=20\nm)=12.6$. (Many authors \cite{DGL11,HYY11} use the intensity ratio for the factor, which corresponds to the square of the FCF in this paper. Even using a much larger tip radius $a$, our result of the efficiency of light concentration is twofold better than the results that have been reported in literature when evaluated by intensity ratio.)

\section{The three underlying effects of the device}
As mentioned above, the field concentration is governed by three effects: mode beat, nanofocusing and weak resonance. We have analyzed in Section~\ref{S:PLC:HybridWG} that the incident mode $\text{TM}_0$ in the bare photonic waveguide can excite two supermodes, $\text{H}_{\text{TM},0}$ and $\text{H}_{\text{TM},1}$, in the corresponding hybrid waveguide with a metal (Au) strip integrated on top. The triangle-shaped PLC can be considered as a taper with the width $w$ of the Au strip gradually going down to $0$. Therefore, we can still use the two supermodes for the Au taper in our analysis. The mode beat between the two supermodes can make a side-coupling which transfers light energy quickly from the dielectric region to the metal region, and the triangle taper performs plasmonic nanofocusing at the same time. A big advantage of our design is that light coupling (induced by mode beat) and focusing are realized in a single triangle-shaped metal taper so that the device can be made ultra-compact. Note that a short taper also means that we cannot separate the processes of supermode excitation, evolution, interference, and coupling between each other. In the following, we will use a simplified model, in which the two supermodes have constant mode effective indices calculated based on the average width of the taper and the exact ratio change of the two supermodes along the taper is not considered.  A certain amount of the light energy will be reflected at the apex and the input end of the taper, and some resonance effect can be observed. This weak resonance also affects the efficiency of how much energy we can concentrate into the tip. By adjusting the size parameters of the Au taper, we can coordinate the three effects and optimize the efficiency of the PLC.
\begin{figure}[htbp]
\centering
\includegraphics[width=.6\textwidth]{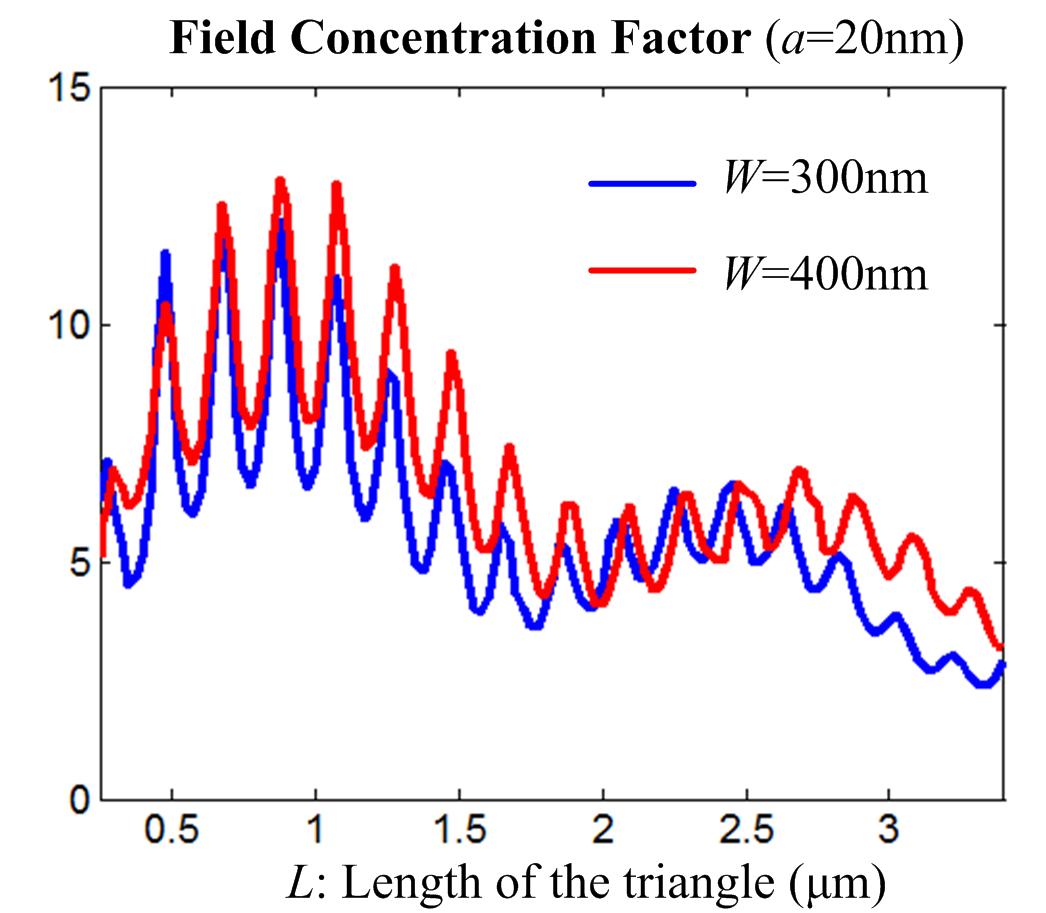}
\caption{FCF vs Length $L$ of the Au triangle.  The width $W$ is $300\nm$ for the blue curve and $400\nm$ for the red curve. The wavelength $\lambda$ is $800\nm$. The radius of curvature $a$ at the tip is $20\nm$.} \label{Fig:PLC:FCF_AuLength}
\end{figure}

Figure~\ref{Fig:PLC:FCF_AuLength} shows how the FCF varies with respect to the length $L$ of the Au triangle, while the radius of curvature $a$ is always set to be $20\nm$. The wavelength $\lambda$ is $800\nm$. The blue curve is for $W=300\nm$, and the red curve if for $W=400\nm$. These curves are composed of two oscillatory parts, the slow variations corresponding to the mode-beat effect and the ripples of high oscillation frequency corresponding to the resonance effect. Both of them can be clearly seen from the figure.

Recall that for a hybrid waveguide, we use Equation~(\ref{Eq:Beatlength}) to estimate the beat length. For the triangle-shaped structure, we may still use Equation~(\ref{Eq:Beatlength}) for beat-length estimation while $w$ is set to be $W/2$ which is about the average width of the triangle. For the blue curve, we have $W=300\nm$ and $w=150\nm$. Therefore, the real part of $n_\text{eff}(\text{H}_{\text{TM},0}(150\nm))$  is $1.9694$, and of $n_\text{eff}(\text{H}_{\text{TM},1}(150\nm))$ is $1.5475$ from Figure~\ref{Fig:PLC:DispersionSiN}. By Equation~(\ref{Eq:Beatlength}), we have
 \begin{equation*}
    \text{BL}(150\nm)=\frac{800\nm}{1.9694-1.5475}=1896\nm.
\end{equation*}
Analyzing analogously on the $W=400\nm$ case (the red curve), we get $w=200\nm$, $\text{Real}(n_\text{eff}(\text{H}_{\text{TM},0}(200\nm)))=1.9216$, $\text{Real}(n_\text{eff}(\text{H}_{\text{TM},1}(200\nm)))=1.5413$, and \\$\text{BL}(200\nm)=2104\nm$. Note that the beat length for the $W=400\nm$ case is a little bit larger than the beat length of the $W=300\nm$ case. The beat lengths can also be estimated from the figure by approximately locating the the first minimums of slow variation parts. These two methods of beat-length estimations agree with each other very well. When $L$ is about half of the beat length, the light energy transfers from the dielectric region to the metal region most efficiently, which results in a higher FCF, as can be seen from the figure. This is actually a principle of design. More accurately speaking, the tapering also induces energy coupling between the two supermodes. In other words, the ratio of energy stored in the two modes will change at different locations along the metal taper, and only the $\text{H}_{\text{TM},0}$ mode becomes localized as the metallic tip.

The ripples on the curves come from the resonance effect. Although the presented structure is not designed based on resonance, the limited length of the taper results in the resonance effect as shown in Figure~\ref{Fig:PLC:FCF_AuLength}. To calculate the free spectral range (FSR), we need to use the effective index of the $\text{H}_{\text{TM},0}$ mode. The formula is
\begin{equation}
    \text{FSR}(w) = \frac{\lambda}{2\text{Real}(n_\text{eff}(\text{H}_{\text{TM},0}(w)))}.
\end{equation}
For example, in the case of $W=400\nm$, we have $w=200\nm$, $\text{Real}(n_\text{eff}(\text{H}_{\text{TM},0}(200\nm)))=1.9216$, and $\text{FSR}(200\nm)=208\nm$. This corresponds to about $15$ peaks for the red curve, a good matching with an actual counting on the figure. The FCF is usually the highest, when $L$ is chosen to be at a resonance peak close to half the beat length. However, we may reduce $L$ to make the PLC supercompact, while still get a high FCF. For example, the FCF is still larger than $10$ with $L=500\nm$ for both curves.

\section{Analysis of the transmission, reflection and FCF spectra}
\begin{figure}[tbp]
\centering
\includegraphics[width=\textwidth]{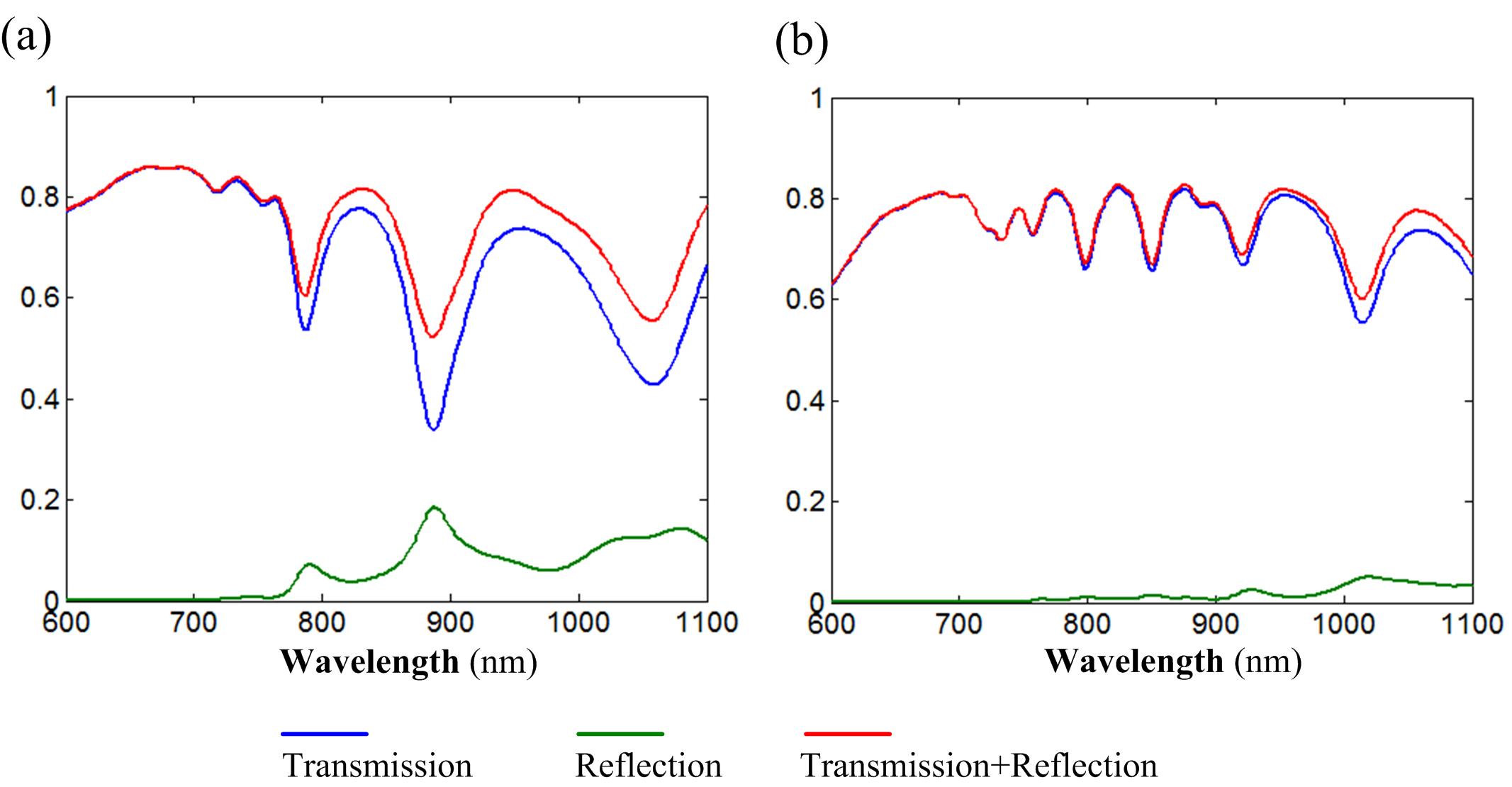}
\caption{Spectra of transmission, reflection and transmission plus reflection for the Au triangle length (a) $L=1\um$ and (b) $L=2\um$. The width $W$ of the triangle is $400\nm$. The radius of curvature $a$ at the tip is $20\nm$.}\label{Fig:PLC:TR1um2um}
\end{figure}

To analyze the transmission spectra, we use the finite-difference-time-domain (FDTD, Lumerical Solutions) method for simulations. Figure~\ref{Fig:PLC:TR1um2um} shows the transmission, reflection and transmission+reflection spectra for (a) $L=1\um$ and (b) $L=2\um$. We have $W=400\nm$ and $a=20\nm$. The wavelength range is $600-1100\nm$. Note that the bare waveguide supports higher order TM modes when the wavelength is less than $700\nm$.

\begin{figure}[htbp]
\centering
\includegraphics[width=.7\textwidth]{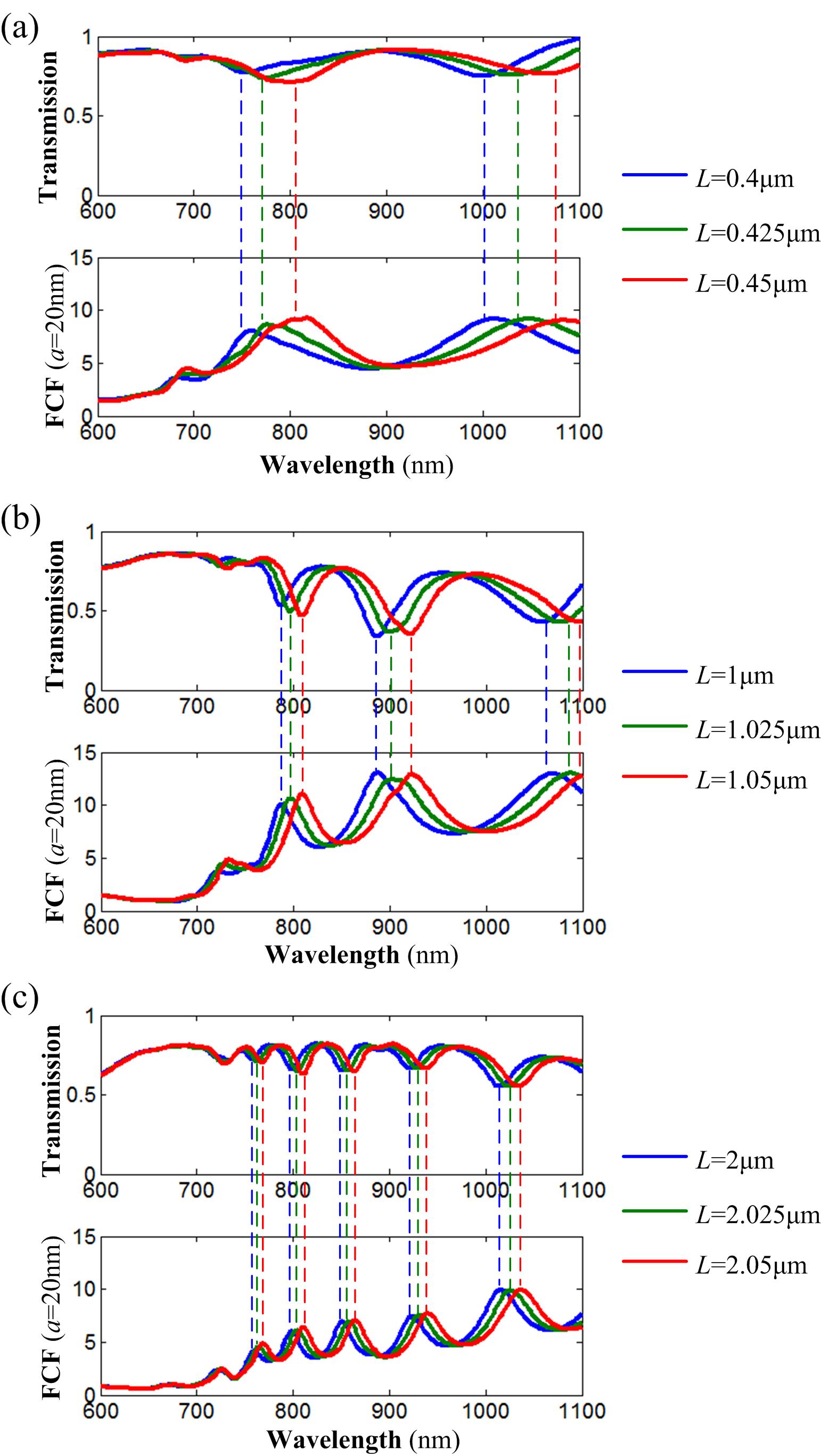}
\caption{Spectra of transmission and FCF for three groups of lengths of Au triangle (a) $0.4\um$, $0.425\um$ and $0.45\um$, (b) $1\um$, $1.025\um$ and $1.05\um$, and (c) $2\um$, $2.025\um$ and $2.05\um$. The width $W$ of the triangle is $400\nm$. The radius of curvature $a$ at the tip is $20\nm$.}\label{Fig:PLC:Transmission_FCF}
\end{figure}

Let us consider the $L=1\um$ case first. As we have computed, when $W=400\nm$, the beat length is about $2.1\um$ at the wavelength of $800\nm$. For wavelengths in a range around $800\nm$, we may assume the beat length is stable. Therefore, $1\um$ is around half the beat length, when the PLC is expected to be most efficient. Let the transmission be $T$, and the reflection be $R$. Then $1-T-R$ indicates the power consumption on the metal structure due to the internal material loss and the radiative loss. As shown in Figure~\ref{Fig:PLC:TR1um2um}(a), the resonance corresponds to the drops in the transmission spectrum and peaks in the reflection spectrum. At resonance, the field is more concentrated in the metal region, which lowers down the transmission but causes a larger reflection. It is noticeable that the reflection part contributes less significantly to the overall $T+R$ than the transmission part, and therefore the profile of the $T+R$ spectrum is basically determined by the transmission spectrum. When the structure is off resonance, we have $T+R$ about $0.8$, which means a certain amount of power is still coupled into the the metal region. This is quite different from a typical spectrum of a (plasmonic) resonator, which can reach a value very close to $1$ when the structure is off resonance. Figure~\ref{Fig:PLC:TR1um2um}(b) shows the spectra in the $L=2\um$ case. The spacing between two successive drops in the transmission spectrum is clearly less than that of the $L=1\um$ case. Moreover, since $2\um$ is close to the beat length, the PLC is expected to be less efficient. As a result, the overall reflection is a lot smaller than in the $L=1\um$ case.


In Figure~\ref{Fig:PLC:Transmission_FCF}, we make comparisons between transmission spectra and FCF spectra for three groups of metal triangle lengths. The first group (Figure~\ref{Fig:PLC:Transmission_FCF}(a)) has lengths $0.4\um$, $0.425\um$ and $0.45\um$, the second group (Figure~\ref{Fig:PLC:Transmission_FCF}(b)) has lengths $1\um$, $1.025\um$ and $1.05\um$, and the third group (Figure~\ref{Fig:PLC:Transmission_FCF}(c)) has lengths $2\um$, $2.025\um$ and $2.05\um$. We still have $W=400\nm$ and $a=20\nm$. Recall that by definition, the FCF($a=20\nm$) is the normalized norm of the electric field at the point $(0,320\nm,20\nm)$ in our coordinate system. When the wavelength is less than $700\nm$, the bare photonic waveguide supports higher order TM modes and the hybrid structure supports more modes other than $\text{H}_{\text{TM},0}$ and $\text{H}_{\text{TM},1}$. Our analytic model using two supermodes is no longer valid in this case. As a result, the field concentration effect is tremendously suppressed. For wavelengths larger than $700\nm$, the vertical dashed lines indicate that a local minima of the transmission corresponds almost exactly to a local maxima of the FCF, which means that the plasmonic structure is at resonance. Note that when $L$ is small, the resonance wavelengths are more sensitive to a change $\Delta L$ of $L$.
\section{Conclusion}
In this paper, we have systematically analyzed the behavior and performance of an ultra-compact plasmonic light concentrator (PLC) which is a triangle-shaped Au taper integrated on a $\text{Si}_3\text{N}_4$ waveguide. We use the fundamental TM-like mode for the incident light as an excitation of two supermodes in the hybrid structure. We have showed that three major effects (mode beat, nanofocusing, and weak resonance) interplay to generate the light concentration phenomenon and govern the performance of the device. The principles of design follow by adequately coordinating these effects. The PLC has highest efficiency when length of the triangle is about half the beat length. In particular, we have demonstrated that after an optimization of the size parameters, a field concentration factor (FCF, defined as the norm of the electric field at the apex normalized to the average norm of the electric field in the inputting waveguide) of about $13$ can be achieved with the length of the device less than $1\um$ for a moderate tip radius $20\nm$. This is about twice better than those being reported in literature with a much larger device size and an assumption of smaller tip radii. In addition, the dimension of the device can be reduced further (a FCF of $5-10$ is achievable with the length of triangle about $400\nm$) when necessary.

Moreover, we point out two more advantages of our device. First, our plasmonic device is mounted on top of a photonic waveguide and the waveguide does not terminate after the light concentration as in typical schemes having been reported in literature \cite{YWLA08,DGL10,DGL11,HYY11}. This makes our device not only simpler in fabrication, but also perfect to fit into the integrated optical circuit and be combined with other units to make a composite components. Second, the three major effects are actually very common and we can use the same philosophy of design for different materials working at different wavelengths.

\section*{Acknowledgement}
The authors would like to thank Ali~A.~Eftekhar for helpful discussions. 
\end{document}